\documentclass[%
 reprint,
superscriptaddress,
groupedaddress,
 amsmath,amssymb,
aps,
pra,
]{revtex4-1}
\usepackage{graphicx}  
\usepackage{dcolumn}   
\usepackage{bm}        
\usepackage{verbatim}  

\usepackage{physics}
\usepackage{float}
\usepackage{color}
\usepackage[dvipsnames]{xcolor}
\usepackage[colorlinks]{hyperref}

\def \dag{^{\dagger}}
\def \ham{\mathcal{H}}
\newcommand{\vect}[1]{\mathbf{\bm{#1}}}
\newcommand{\uvec}[1]{\hat{\vect{#1}}}
\newcommand{\vsigma}{\vect{\sigma}}

\begin{document}

\title{Synergetic effect of spin-orbit coupling and Zeeman splitting on the optical conductivity in the one-dimensional Hubbard model}
\author{Adrien Bolens, Hosho Katsura, Masao Ogata, Seiji Miyashita}
\affiliation{
    Department of Physics, University of Tokyo, Hongo, Bunkyo-ku, Tokyo 113-0033, Japan
   }

\date{\today}
\begin{abstract}
We study how the synergetic effect of spin-orbit coupling (SOC) and Zeeman splitting (ZS) affects the optical conductivity in the one-dimensional Hubbard model using the Kubo formula. We focus on two phenomena: (1) the electric dipole spin resonance (EDSR) in the metallic regime and (2) the optical conductivity in the Mott-insulating phase above the optical gap. In both cases, we calculate qualitatively the effects of SOC and ZS and how they depend on the relative angle between the SOC vector and the magnetic field direction.
First, we investigate the spin resonance without electron correlation (the Hubbard parameter $U=0$). Although, neither SOC nor ZS causes any resonance by itself in the optical conductivity, the EDSR becomes possible when both of them exist. The resulting contribution to the optical conductivity is analyzed analytically. The effect of $U$ on the spin resonance is also studied with a numerical method. It is found that at half-filling, the resonance is first enhanced for small $U$ and then suppressed when the optical gap is large enough.
In the strong coupling limit $U \rightarrow \infty$ at half-filling, we also refer to the resonance between the lower and upper Hubbard bands appearing at $\omega \sim U$, above the optical gap. A large magnetic field tends to suppress the signal while it is recovered thanks to SOC depending on the relative angle of the magnetic field. 
\end{abstract}

\maketitle

\section{Introduction}

Measurements of dynamical response functions are expected to test the validity of existing theories of one-dimensional (1D) systems. The dynamical properties in the metallic regime are often well understood in terms of the Tomonaga-Luttinger liquid theory. The Mott-insulating phase \cite{mott1990metal}, however, is best described by the Hubbard model, which also incorporate the effect of the lattice potential.

In this paper, we are interested in the synergetic effect of spin-orbit coupling (SOC) and Zeeman splitting (ZS) on the dynamical response functions of the 1D Hubbard model, in particular the optical conductivity. We aim to study how spin dynamics affects the optical conductivity through SOC. We focus on two phenomena: (1) the electron spin resonance (ESR) in the itinerant regime and (2) the optical conductivity in the Mott-insulating phase above the optical gap \cite{lieb1968absence, ovchinnikov1992excitation} in the strong coupling limit.

ESR measures the absorption of an electromagnetic (EM) wave by electrons in a constant magnetic field and is a useful tool to study the dynamics of electron spins. In a metal with an initial spin-SU($2$) symmetry, the constant magnetic field breaks the SU($2$) symmetry and the ESR corresponds to the absorption of the magnetic component of the EM wave and has a $\delta$ function peak at the energy corresponding the ZS (even with interaction \cite{oshikawa2002electron}). However, when the SOC exists, ESR is also possible through the electric component of the EM wave. In this case, the absorption rate of ESR is proportional to the optical conductivity, as shown shortly. Indeed, the SOC creates an effective magnetic field that is proportional to the electron momentum. This is called the electric dipole spin resonance (EDSR) \cite{rashba1960properties, rashba1991electric, rashba2003orbital, efros2006theory}.

In a non-interacting one-band 1D electron system without SOC (e.g. tight-binding model), the optical conductivity has only a Drude weight contribution at $\omega = 0$ and vanishes for $\omega > 0$, $\sigma(\omega) = D\delta(\omega)$. In the presence of SOC, the only contribution to $\sigma(\omega>0)$ is the EDSR.
Other contributions at $\omega > 0$ are typically due to umklapp process \cite{giamarchi1991umklapp} once interaction are introduced. 

Therefore, in a metal with SOC, an ESR experiment measures both the optical conductivity and the spin susceptibility at the spin resonance frequency $\omega_{\rm res}$. The absorption rate of an incident EM wave with frequency $\omega_{\rm res}$ and constant amplitude $\tilde{E}_0$ (along the 1D system) and $\tilde{B}_0$ is (in the linear response theory)
\begin{equation}
\label{eq:abs}
  I(\omega_{\rm res}) = 2 \sigma'(\omega_{\rm res}) \tilde{E}_0^2+ 2 \omega_{\rm res} (g \mu_B)^2 \chi''(\omega_{\rm res}) {\tilde B}_0^2,
\end{equation}
where $\sigma'(\omega)$ is the real part of the optical conductivity, $\chi''(\omega)$ is the imaginary part of the spin susceptibility (along the direction of the ac magnetic field), $g$ is the Land\'e factor and $\mu_B$ is the Bohr magneton.
Due to the relative strength of the magnetic dipole contribution and the electric dipole contribution, the magnetic dipole ESR amplitude (the second term in Eq.~(\ref{eq:abs})) is much weaker than the EDSR amplitude (the first term in Eq.~(\ref{eq:abs})) by several orders of magnitude \cite{shekhter2005chiral,maiti2016electron}. 
Indeed, when SOC is of the same order as the ZS, the relative contributions are of the order $(a/\lambdabar_{\rm C})^2$ where $a$ is the lattice spacing (typically, $a \approx 5\times 10^{-10}$ m) and $\lambdabar_{\rm C} = \hbar/m_{\rm e} c$ is the reduced Compton length of the electron ($\approx 3.86 \times 10^{-13}$ m).
 
In the first studies of EDSR in a two-dimensional (2D) electron gas \cite{rashba1960properties, rashba1991electric, rashba2003orbital, efros2006theory}, the ZS was considered to be much larger than the magnitude of the SOC. Thus, SOC was treated perturbatively up to first order. In this case, the EDSR signal gives a $\delta$ function peak at  frequency of the ZS, similarly to the usual paramagnetic resonance. On the other hand, when SOC cannot be treated perturbatively, a resonance appears even without an external magnetic field due to the SOC splitting \cite{magarill2001spin, mishchenko2003transport, shekhter2005chiral, farid2006optical}. Indeed, the SOC breaks the SU($2$) symmetry and splits the free-electron dispersion in two branches corresponding to different chiralities, and an electric dipole transition is allowed between the two branches. In this case, the absorption spectrum has a finite width (boxlike shape) around frequencies corresponding to the SOC splitting at the Fermi surface.
Studies on response function with SOC have been discussed from various viewpoints. For example, the effect of Coulomb interaction on collective modes in the system with SOC has been studied \cite{maiti2015collective}.
The synergetic effect of ZS and SOC on the optical conductivity has been investigated only recently. In the free 2D electron gas, the shape of the EDSR amplitude has been calculated \cite{glenn2012interplay} and the effect of SOC and magnetic field on the collective modes has been studied in a Fermi liquid \cite{maiti2016electron}.

However, in one dimension, EDSR is only possible when a static magnetic field is applied with a finite component perpendicular to the internal magnetic field due to SOC, without which the EDSR is not allowed \cite{abanov2012spin}. The effects of the interplay of SOC and magnetic field on the spin resonance has been studied in a free quantum wire \cite{abanov2012spin} and in a Tomonaga-Luttinger liquid \cite{tretiakov2013spin, gangadharaiah2008spin}.

In the present paper, we study the effect of a lattice potential on the EDSR in the 1D tight-binding (TB) model. The TB model is the simplest model which can capture the effect of the lattice potential and has a cosine-like dispersion relation. More importantly, the SOC Hamiltonian is also changed and adopts a periodic shape with the periodicity of the Brillouin zone. Therefore, the resulting effect of SOC strongly depends on the position of the Fermi energy in the band. In the limit of quadratic dispersion, which corresponds to a Fermi energy lying in the very bottom or very top of the band, the results agree with those in an electron gas \cite{glenn2012interplay,abanov2012spin}.

Without violation of the translational symmetry and without SOC, the gauge field couples directly to the center of mass of the system, and the dynamics of the total current is unaffected by interactions \cite{shekhter2005chiral}. 
However, in our case, this is no longer true and the time evolution of the current depends on the interaction. We therefore also study the effect of an on-site Hubbard interaction numerically, using exact diagonalization.

Finally, we also refer to the optical conductivity in the $U \rightarrow \infty$ limit of the Hubbard model at half-filling. In this case, the charge degree of freedom is completely gapped and the system is a Mott insulator for any $U>0$ \cite{lieb1968absence}. The EDSR is no longer observed in this limit. We study the synergetic effects of SOC and magnetic field on the optical resonances between the two Hubbard bands, separated by the optical gap $\Delta_{\rm opt}$, at high frequencies of the order $\omega \sim U$.

The rest of the paper is organized as follows. In section \ref{sec:model}, we define the model containing both SOC and the ZS, and define the quantities which are necessary for the calculation of the optical conductivity. In section \ref{sec:OC}, we calculate the optical conductivity in the TB model and in subsection \ref{sec:OCHubbard} we analyze the effect of the Hubbard interaction and show exact numerical results for small systems. In section \ref{sec:largeU} we investigate the model in the large $U$ limit and we finally conclude in section \ref{sec:conclusion}.

\section{Models}
\label{sec:model}
\subsection{1D Tight-Binding Model\label{sec:TB}}

We consider a model of interacting itinerant electrons in a 1D crystal with SOC in an external magnetic field. SOC is due to inversion asymmetry of the system, i.e. systems with Rashba SOC caused by the structural asymmetry \cite{bychkov1984properties}, and systems with Dresselhaus SOC caused by bulk asymmetry \cite{dresselhaus1955spin}. As a model which can capture the above mentioned physics we consider, 

\begin{align}
  \label{eq:tight}
    \mathcal{H} =& -t\sum_{l}( {\bm c}\dag_{l+1} \sigma_0 {\bm c}_{l} + \text{H.c.})  - \sum_{l} {\bm c}\dag_{l} (\vect{b} \cdot {\bm \sigma} ) {\bm c}_l\nonumber \\
    &+ \mathcal{H}_{\rm SO}  + U\sum_{l} n_{l \uparrow} n_{l \downarrow},\\
     \mathcal{H}_{\rm SO} =& i\lambda \sum_{l} ({\bm c}\dag_{l+1}(\uvec{d} \cdot \vsigma ) {\bm c}_{l} - \text{H.c.}),
    \label{eq:SOCham}
\end{align}
where $t$ is the transfer integral and ${\bm c}_l\dag = (c_{l\uparrow} \dag, c_{l\downarrow} \dag)$, such that $c\dag_{ls}$($c_{ls}$) is the electron creation (annihilation) operator at the lattice site $l$ with spin $s$. Here $(\sigma_0,{\bm \sigma})$ denote the identity and the Pauli matrices, respectively, $\vect{b} = (g\mu_{B}/2)\vect{B}$ where $\vect{B}$ is the external magnetic field, and $U$ is the strength of the Hubbard interaction. Up to section \ref{sec:OCTB}, we discuss the non-interacting limit and we set $U=0$. In 1D, the orbital effects of $\vect{B}$ can always be gauged away. 
$\ham_{\rm SO}$ denotes the effective SOC Hamiltonian in the lattice model due to the inversion asymmetry, where $\lambda$ is the SOC parameter and $\uvec{d}$ is the normalized SOC vector (i.e. the direction of the SOC internal magnetic field), whose direction is assumed constant throughout the system. Once casted onto the Wannier functions, the SOC makes the electrons rotate their spin when they hop between neighbouring sites. Therefore, in a lattice system, the SOC Hamiltonian can be understood as an SU($2$) gauge field \cite{hatano2007non}. The electrons acquire a spin dependent phase, the Aharonov-Casher phase \cite{aharonov1984topological}. The resulting SOC Hamiltonian is given by (\ref{eq:SOCham}). The parameters $\lambda$ satisfies $\lambda/t=\tan(\theta/2)$, where $\theta$ is the angle the spin of an electron rotates by whenever the electron hops between two sites.

The Hamiltonian (\ref{eq:tight}) with $U=0$ can be rewritten by performing a Fourier transformation as
\begin{equation}
  \ham = \sum_{k} \vect{c}\dag_{k} \ham(k) \vect{c}_{k},
\end{equation}
with
\begin{align} \label{eq:TBham}
    \ham(k) &= -2t \cos(k) \sigma_0 + \left( 2\lambda \sin (k) \uvec{d}  - \vect{b} \right) \cdot \vsigma \nonumber \\
    &= -2t \cos(k) \sigma_0 + \Delta(k) \uvec{n}_k \cdot \vsigma,
\end{align}
where we set the lattice spacing $a=1$.
$k$ is the crystal quasi-momentum of the Bloch wave-function. ${\bm c}_k \dag = (c_{k \uparrow} \dag, c_{k \downarrow} \dag)$, where $c\dag_{k s}$($c_{k s}$) is the electron creation (annihilation) operator with momentum $k$ and spin $s$. $\Delta(k)$ and $\uvec{n}_{k}$ are defined as
\begin{align}
  \Delta(k) &= \norm{2\lambda \sin(k) \uvec{d} - \vect{b}}, \label{eq:Delta} \\
  \uvec{n}_{k} &= \frac{2\lambda \sin(k) \uvec{d} - \vect{b}}{\Delta(k)},
\end{align}
such that $\uvec{n}_{k}$ has unit length and $\Delta(k)$ can be thought of as the effective Zeeman splitting for the wave vector $k$. Let us additionally define $\varphi$ as the angle between $\vect{b}$ and $\uvec{d}$ and let us decompose $\vect{b} = \vect{b}_{\perp} + \vect{b}_{\parallel}$ so that $\vect{b}_{\parallel}$ is along $\uvec{d}$ and $\vect{b} \cdot \uvec{d} = \vect{b}_{\parallel} \cdot \uvec{d} = b \cos(\varphi)$, where $b$ is the magnitude of $\vect{b}$.

Then, using an SU($2$) gauge transformation on the spin basis, the energy dispersion is split in two branches corresponding to spin up and down in the $\uvec{n}_{k}$ direction.
The energy branches are separated by $2\Delta(k)$ and are given by
\begin{equation} \label{eq:energy}
  \varepsilon^{\pm}(k) = \varepsilon_0(k) \pm \Delta(k).
\end{equation}

\begin{figure}[htb]
\centering
	\includegraphics[width=0.4\textwidth]{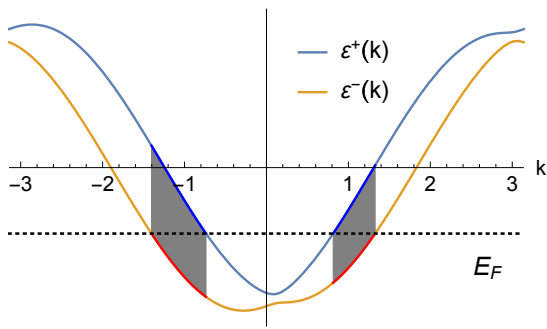}
	\caption{Dispersion relation in the 1D tight-binding model. The gap between the two lines is $2\Delta(k)$. The shaded regions correspond to momenta contributing to the optical conductivity at $T=0$ at frequencies $\omega = 2\Delta(k)$ (see Eq.~\ref{eq:1Dcond}). The parameters are $t=1$, $\lambda = 0.3$, $b=0.1$ and $\varphi = \pi/4$.}
\label{fig:1d}
\end{figure}

As an example, Fig.~\ref{fig:1d} shows the dispersion relation with parameters chosen such that the effects of SOC and ZS on the dispersion are very clear ($t=1$, $\lambda = 0.3$, $b=0.1$, $\varphi = \pi/4$). 

The unitary transformation $U_{k}$ that diagonalizes $\mathcal{H}(k)$ is defined by 
\begin{equation}
  U_{k}\dag( \uvec{n}_{k} \cdot {\bm \sigma} )U_{k} = \sigma_z,
\end{equation}
and the eigen-operator ${\bm {\tilde c}}_{k} = (\tilde c_{k+}, \tilde c_{k-})$, which annihilates a fermion with energy given by (\ref{eq:energy}), is
\begin{align}
   {\bm {\tilde c}}_{k} = U_{k}\dag {\bm c}_{k}.
\end{align}

Let us now define the current operator for the tight-binding model, which will be used to calculate the conductivity. Due to the discrete nature of the system, the definition of current is not obvious. A physically satisfactory definition (see e.g. Ref.~\cite{mahan2013many}) relies on the polarization operator defined on the lattice as
\begin{equation}
\label{eq:polar}	
  P = e \sum_{l} x_l n_{l},
\end{equation}
where $x_l$ is the position of the lattice site $l$ and $e$ is the elementary charge. The total current operator $j$ is defined as
\begin{equation}
  \label{eq:current}
  j \equiv \dot{P} = -i[P,\ham].
\end{equation}
One can check that such a definition is consistent with the discrete continuity equation of the lattice system. 

Plugging the Hamiltonian (\ref{eq:tight}) into Eq.~(\ref{eq:current}), we obtain
\begin{align}
  j =&  e\sum_{l} {\bm c}\dag_{l + 1} (it\sigma_0 + \lambda \uvec{d} \cdot {\bm \sigma} )  {\bm c}_{l} + \text{H.c.} \nonumber \\
   =&\sum_{k} {\bm c}\dag_{k} j(k) {\bm c}_{k},
\end{align}
where $j(k)$ is analogous to the group velocity of the electrons ($j(k) = e\partial_k \ham(k)$) and is given by

\begin{align}
  j(k) = & 2et\sin(k) \sigma_0 + 2e\lambda \cos(k) \uvec{d} \cdot {\bm \sigma} \nonumber \\
  \equiv & j_{\rm K}(k) \sigma_0 + j_{\rm SO}(k)  \uvec{d}\cdot \vsigma.
\end{align}

Only the spin-orbit induced current ($j_{\rm SO}$) gives rise to the EDSR signal, while both the ``kinetic'' current ($j_{\rm K}$) and the SO current contribute to the Drude part of the conductivity.
In terms of energy eigenstates, the current operator reads
\begin{equation}
\label{eq:newcur}
   U_{k}\dag j(k) U_{k} = j_{\rm K} (k) \sigma_0 + j_{\rm SO}(k)  \uvec{d}_k' \cdot \vsigma,
\end{equation}
where $\uvec{d}_k' =  U_{k}\dag \uvec{d} U_{k}$. The transformation defined by $U_{k}$ correspond to an SO($3$) rotation in $\mathbb{R}^3$ which preserves angles. Therefore, we have the relation
\begin{equation}
\label{eq:angle}
  \uvec{d} \cdot \uvec{n}_{k} = \uvec{d}_k' \cdot \uvec{z}.
\end{equation}

\section{Optical conductivity}
\label{sec:OC}
Let a uniform ac electric field be polarized along the $x$ direction,
\begin{equation}
  \tilde{\vect{E}}(t) = \tilde{E}_0 \cos(\omega t) \uvec{e}_x.
\end{equation}
In a lattice model, it is most convenient to introduce the coupling to the field through a scalar potential $V(x) = -E x$. The total Hamiltonian is then $\ham' = \ham + E P$, where $P$ is the polarization defined in (\ref{eq:polar}).
In the linear response theory, the real part of the optical conductivity at positive frequency $\omega$ is given by the Kubo formula as
\begin{equation}
\label{eq:OCgeneral}
  \sigma'(\omega) = \frac{1}{L \omega} {\rm Re} \left\{ \int_{0}^{\infty}e^{i\omega t}\langle \left[ j(t), j(0) \right] \rangle dt \right\},
\end{equation}
where $L$ is the length of the system.
For non-interacting particles, the linear response can be written in terms of single-particle states $\ket{k \pm} = \tilde{c}_{k, \pm}\dag\ket{0}$, where $\ket{0}$ is the vacuum state (i.e. with no electrons):
\begin{align}
  \sigma'_{i}(\omega) = \frac{\pi}{L\omega} \sum_{k} &\abs{\mel{k +}{j}{k -}}^2 \delta(\omega - 2\Delta(k)) \nonumber \\
  &\times \left( \langle n_{k -} \rangle - \langle n_{k +} \rangle \right),
  \label{eq:opcond}
\end{align}
where $\langle \cdot \rangle$ denotes the grand canonical ensemble average. Therefore $\langle n_{k \pm} \rangle = n_F(\varepsilon^{\pm}(k))$, where $n_F$ is the Fermi-Dirac distribution.
With Eq.~(\ref{eq:opcond}), we can calculate the optical conductivity due to SOC of any free one-band model with translational invariance (continuous or discrete).
\\

The SI unit of optical conductivity is the siemens per meter ($S/m$). In a 1D system, however, the current density has units of Ampere and thus the 1D optical conductivity (as defined in Eq.~\ref{eq:OCgeneral}) has units of simens-meter ($S \cdot m$). From now on, we set $e=1$, together with $a$ and $\hbar$. Within this convention, $\sigma'(\omega)$ is unitless. In particular, it is invariant under a rescaling of the energy. $\sigma'(\omega)$ has to be multiplied by $e^2 a / \hbar$ to recover the proper units in all the following expressions and graphs. The standard (3D) optical conductivity can be obtained using $\sigma_{3{\rm D}} = \sigma_{1 {\rm D}}/a_{\perp}^2$, where $a_{\perp}$ is the lattice spacing perpendicular to the chain direction.

\subsection{Optical conductivity of the 1D tight-binding model}
\label{sec:OCTB}

Using Eqs. (\ref{eq:newcur}) and (\ref{eq:angle}), off-diagonal matrix elements for the current are given by
\begin{align}
  \abs{\mel{k +}{j}{k -}}^2 &= {j_{\rm SO}(k)}^2\left[ 1- (\uvec{d}_k' \cdot \uvec{z})^2 \right] \nonumber \\
  &= {j_{\rm SO}(k)}^2 \left[ 1- (\uvec{d} \cdot \uvec{n}_k)^2 \right].
  \label{eq:mel}
\end{align}

The optical conductivity then reads
\begin{align}
  \label{eq:1Dcond}
  \sigma'(\omega) =& \frac{1}{L} \frac{16 \pi }{\omega^3} \lambda^2 b_{\perp}^2  \\
  &\times \sum_{k} \cos(k) \left( \langle n_{k -} \rangle - \langle n_{k +} \rangle \right) \delta(\omega - 2\Delta(k)). \nonumber
\end{align}

Note that $\sigma'(\omega)$ depends on $\varphi$ through both $\vect{b}_{\perp}$ and $\Delta(k)$.
In one dimension, the Fermi surface is just two points, each of which is in the $+$ or $-$ branch corresponding to left and right moving electrons. At $T=0$, only momenta between the two Fermi surfaces, indicated by the shaded regions in Fig.~\ref{fig:1d}, contribute to the conductivity.

\begin{figure}[htb]
\centering
	\includegraphics[width=0.43\textwidth]{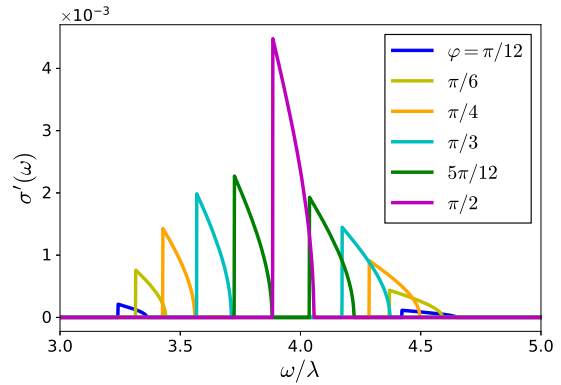}
	\caption{Optical conductivity of the 1D tight-binding model without the interaction $U$ at half-filling, for different orientations of the magnetic field at $T=0$. The optical conductivity was calculated using Eq.~(\ref{eq:1Dcond}) and is unitless, as explained in the main text. The parameters are chosen such that the effects are enhanced ($t=1$, $\lambda = 0.3$, $b=0.1$).}
\label{fig:OClargepara}
\end{figure}

Figure \ref{fig:OClargepara} shows the optical conductivity at half-filling ($E_F = 0$) calculated with the same parameters as in Fig.~\ref{fig:1d} ($t=1$, $\lambda = 0.3$, $b=0.1$), for different values of $\varphi$. 
The optical conductivity has two peaks around $\omega = 2\Delta(\pm k_F)$, where $-2t\cos(k_F) = E_F$, except for the case with $\varphi = \pi/2$. Indeed, the two peaks merge together when $\vect{b}$ and $\uvec{d}$ are perpendicular as the $-k \leftrightarrow k$ symmetry is recovered. The peaks have an intrinsic width even at $T=0$, as we see in Fig.~\ref{fig:OClargepara}. 

\begin{figure}[ht!]
\centering
\includegraphics[width=0.46\textwidth]{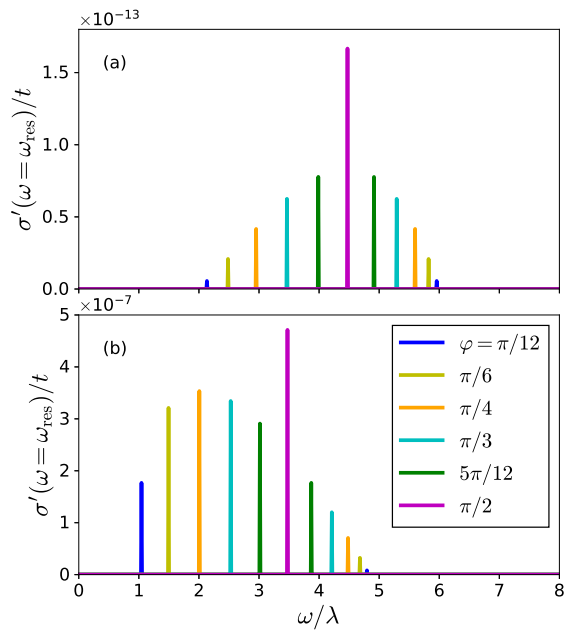}
\caption{Optical conductivity of the 1D tight-binding model without the interaction $U$ for $\lambda/t = b/t = 10^{-3}$ at half-filling (a) and quarter-filling (b) for different orientations of the magnetic field at $T=0$. The optical conductivity was calculated with Eq.~(\ref{eq:1Dcond}) and is plotted in units of $t$ since the dirac delta function has been factored out, as explained in the main text.}
\label{fig:OCrealpara}
\end{figure}

We then consider the realistic limit where $t \gg \lambda \sim b$. In our calculations, we set $\lambda/t = b/t = 10^{-3}$, which, for $t=100$ meV, corresponds to a magnetic field of around $1.7$ T. Figure \ref{fig:OCrealpara} shows the optical conductivity at half- and quarter-filling. With those parameters, the peaks are narrow and all contributing momenta $k$ in Eq.~(\ref{eq:1Dcond}) correspond to the same frequency $\omega_{\rm res} = 2\Delta(\pm k_F)$ up to reasonable numerical resolution. Therefore, we can factorize $\delta(\omega - \omega_{\rm res})$ and we only show the multiplying factor, which is plotted in units of $t$.

The magnitude of $\sigma'(\omega)$ depends on $E_F$ due to the $\cos(k)$ term in (\ref{eq:1Dcond}), notably at $E_F = 0$ (half-filled) where the signal is highly suppressed due to the small spin-orbit current $j_{\rm SO}(k) \propto \cos(k)$.
The optical conductivity is only finite when both $b_{\perp}$ and $\lambda$ are non-zero as $\sigma'(\omega)$ is proportional to $\lambda^2 b_{\perp}^2$. This is also the case in the corresponding continuum model \cite{abanov2012spin}.
This is in contrast to the 2D case where SOC allows a spin resonance by electric field even when $b=0$ \cite{magarill2001spin, mishchenko2003transport, shekhter2005chiral, farid2006optical}.

\subsection{Interaction and Hubbard Model\label{sec:OCHubbard}}
We now study the effect of the Hubbard interaction:t
\begin{equation}
\label{eq:U}
  \ham_{U} = U\sum_{l} n_{l \uparrow} n_{l \downarrow}.
\end{equation}
Here, we investigate the effect of interaction on the spin resonance using exact diagonalization, in small systems. We study the system in two situations corresponding to half-filling ($7$ electrons in $7$ sites) and quarter-filling ($4$ electrons in $8$ sites). In the canonical ensemble, the optical conductivity for $\omega>0$ is calculated from Eq.~(\ref{eq:OCgeneral}) and is given by
\begin{equation}
  \sigma'(\omega) = \frac{\pi}{L}\underset{E_m \neq E_n}{\sum_{mn}}\frac{e^{-\beta E_m}-e^{-\beta E_n}}{Z}\frac{\abs{\mel{\psi_m}{j}{\psi_n}}^2}{\omega_{mn}} \delta(\omega - \omega_{mn}),
\end{equation}
where $\ket{\psi_m}$ is an eigenstate with energy $E_m$, $\omega_{mn} = E_n - E_m$, $Z$ is the partition function and $\beta = 1/k_BT$ is the inverse temperature. In the following, we only discuss the $T=0$ situation.
 
The 1D Hubbard model was solved exactly by Lieb and Wu \cite{lieb1968absence}. They showed that at half-filling, the system is an insulator with a finite optical gap $\Delta_{\rm opt}$, i.e. a gap in the optical absorption, for any positive $U \neq 0$. Moreover, away from half-filling, the system is metallic with no optical gap for all $U$.
In our case, $U$ has two effects on the optical conductivity $\sigma'(\omega>0)$: (1) it modifies the EDSR contribution which is studied at $U=0$, (2) it enables resonances from the ground state to the new optically allowed states. In this section, we only consider the effect (1), while the effect (2) is discussed in the next section, in the $U/t \rightarrow \infty$ limit.

The size of the optical gap has been calculated exactly \cite{ovchinnikov1992excitation} and approaches $\Delta_{\rm opt} \propto \sqrt{U/t}\exp{-2\pi t/U}$ as $U \rightarrow 0$, which means that the gap is exponentially suppressed when $U \lesssim 2t$. Those results are still valid with the addition of SOC as its effect can be gauged away (although corrections are needed for a finite size system due to the periodic boundary conditions \cite{fujimoto1993persistent}), but not with the addition of a magnetic field. However, we do not expect a qualitative change as $b$ is small compared to $t$.

\begin{figure}[ht!]
\centering
\includegraphics[width=0.4\textwidth]{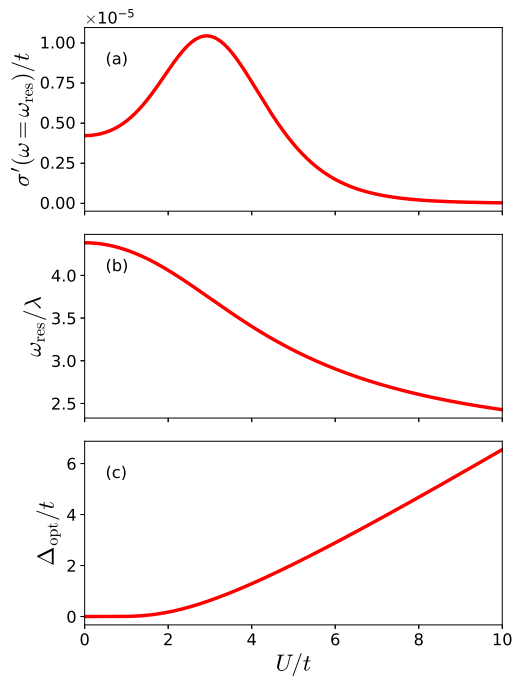}
\caption{(a) The optical conductivity at $\omega_{\rm res}$, and (b) $\omega_{\rm res}$ plotted as a function of the interaction strength $U$ for $\varphi = \pi/2$ ($b_{\parallel} = 0$) at half-filling ($7$ electrons in $7$ sites) with parameters $\lambda/t = b/t = 10^{-3}$. (c) The exact optical gap in the thermodynamic limit (without magnetic field) as calculated in Ref.~\cite{lieb1968absence, ovchinnikov1992excitation}.}
\label{fig:oc1D}
\end{figure}

Figure \ref{fig:oc1D} shows the results for the half-filling case with $\varphi = \pi/2$ (i.e. $b_{\parallel} = 0$) as well as the exact optical gap $\Delta_{\rm opt}$ in the thermodynamic limit calculated using the results from Ref.~\cite{lieb1968absence, ovchinnikov1992excitation}. The parameters are set to $\lambda/t = b/t = 10^{-3}$ so that the peak in the response function due to the spin resonance is very narrow (see Fig.~\ref{fig:OCrealpara}). Therefore, we characterize the spin resonance only by the peak amplitude and its frequency $\omega_{\rm res}$. The amplitude of the spin resonance existing at $U=0$ varies continuously and a shift in the frequency $\omega_{\rm res}$ is observed, as shown in Fig.~\ref{fig:oc1D}~(a) and (b). It is found that for small $U$, the amplitude is first enhanced. But, as $U$ gets larger, it is suppressed as the optical gap grows and reaches $\Delta_{\rm opt} \sim t$, where we naturally expect localized spins to appear. 

\begin{figure}[ht!]
\centering
\includegraphics[width=0.4\textwidth]{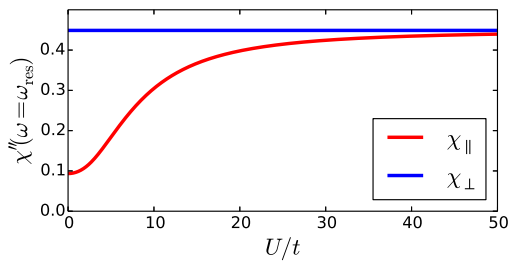}
\caption{The magnetic susceptibility at $\omega_{\rm res}$ is plotted as a function of the interaction strength $U$ for $\varphi = \pi/2$ ($b_{\parallel} = 0$) at half-filling ($7$ electrons in $7$ sites) with parameters $\lambda/t = b/t = 10^{-3}$. The ac magnetic field is chosen perpendicular to $\vect{b}$ and either parallel ($\chi_{\parallel}$) or perpendicular ($\chi_{\perp}$) to $\uvec{d}$.}
\label{fig:ms1D}
\end{figure}

The dynamical spin susceptibility at the spin resonance is shown in Fig.~\ref{fig:ms1D} as a function of $U$. The polarization of the incident ac magnetic field $\tilde{\vect{B}}(t)$ is chosen perpendicular to the static magnetic field $\vect{b}$ and either perpendicular or parallel to the SOC vector $\uvec{d}$. The corresponding magnetic susceptibility is denoted by $\chi_{\perp}(\omega)$ or $\chi_{\parallel}(\omega)$, respectively. In the case of $\tilde{\vect{B}} \perp \uvec{d}$, the effective Zeeman field in Eq.~(\ref{eq:TBham}), $\Delta(k)\uvec{n}_k$, is always perpendicular to the ac magnetic field so that the direction of $\uvec{n}_k$ does not affect the spin resonance, and the amplitude of $\chi_{\perp}$ is unaffected by SOC and thus the interaction \cite{oshikawa2002electron}, as can be seen in Fig.~\ref{fig:ms1D}. In the case of $\tilde{\vect{B}} \parallel \uvec{d}$, $\uvec{n}_k$ is not perpendicular to $\tilde{\vect{B}}$ and the SOC affects the amplitude of $\chi_{\perp}$, so that $\chi''_{\perp}(\omega_{\rm res})$ increases with $U$ until it reaches the value it would have without SOC. Indeed, for large $U$ it corresponds to the spin susceptibility of the effective localized spin model discussed in the next section, in which the effect of SOC scales as $t\lambda/U$. 
In any case, the interplay of SOC, ZS and the interaction causes a shift in the frequency $\omega_{\rm res}$ of the spin resonance, as shown in Fig.~\ref{fig:oc1D}~(b).

\begin{figure}[ht!]
\centering
\includegraphics[width=0.4\textwidth]{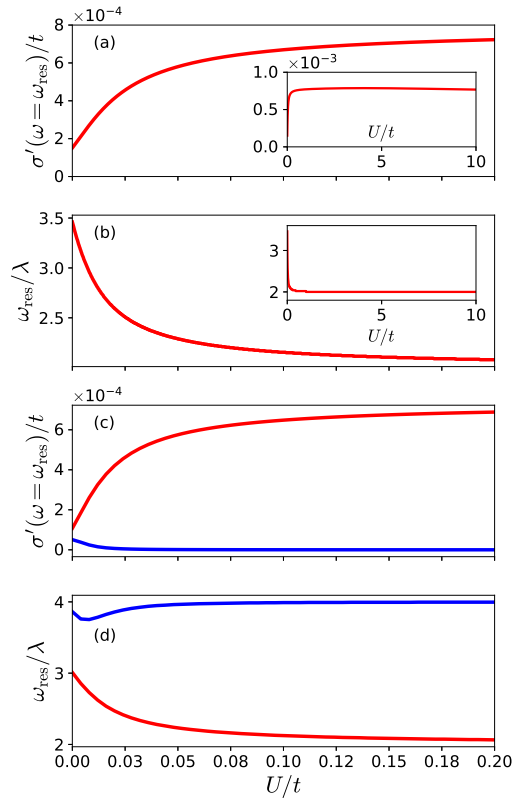}
\caption{The optical conductivity at $\omega_{\rm res}$ and $\omega_{\rm res}$ are plotted as a function of the interaction strength $U$ at quarter-filling ($4$ electrons in $8$ sites) for two different values of $\varphi$ with parameters $\lambda/t = b/t = 10^{-3}$. (a) and (b): The amplitude and the resonance frequency for $\varphi = \pi/2$ ($b_{\parallel} = 0$) are depicted, respectively. The insets represent the same plots in an extended region of $U$. (c) and (d): The amplitudes and the resonance frequencies for $\varphi = 5\pi/12$ ($b_{\parallel} \neq 0$) are depicted, respectively.}
\label{fig:oc1D_quarter}
\end{figure}

Figure \ref{fig:oc1D_quarter} shows the results for the quarter-filling case with $\varphi = \pi/2$ ($b_{\parallel} = 0$) and $\varphi = 5\pi/12$ ($b_{\parallel} \neq 0$). For $\varphi = \pi/2$, for small values of $U/t$, the interaction modifies the amplitude and shifts the frequency $\omega_{\rm res}$. The amplitude converges as $U$ increases, so that the EDSR is not destroyed in the strongly correlated regime. This is in agreement with the results of Ref.~\cite{tretiakov2013spin} in the Tomonaga-Luttinger liquid. We therefore infer that the EDSR signal is suppressed by interaction only in the insulating phase with a finite optical gap.
For $\varphi = 5\pi/12$, the spin resonance of the left and right moving electrons occurs at different frequencies as $b_{\parallel} \ne 0$. We thus observe the evolution of two peaks in the optical conductivity, as shown in Fig.~\ref{fig:oc1D_quarter}~(c) and (d). As with $\varphi = \pi/2$, the interaction modifies the amplitude and shifts the resonant frequencies. Interestingly, only one peak (with the smaller resonant frequency) survives at large $U$. 
\section{Large coupling limit}
\label{sec:largeU}
Finally, we refer to the large coupling limit $U/t \rightarrow \infty$ of the Mott-insulating phase at half-filling.
The optical conductivity of the Hubbard model has been extensively studied with both analytic and numerical methods at half-filling \cite{jeckelmann2000optical,jeckelmann2002dynamical,controzzi2001optical,essler2001excitons} and away from half-filling \cite{giamarchi1991umklapp,giamarchi1992conductivity,giamarchi1997mott,horsch1993frequency,tiegel2016optical}. 
 
Note that the EDSR contribution to the optical conductivity discussed in the previous section vanishes in the large coupling limit (see Fig.~\ref{fig:oc1D}). Thus, this contribution is not discussed in this section and we study the effect of SOC and magnetic field on the optical conductivity above the optical gap. 

In the large $U$ limit, without SOC nor magnetic field, the charge degree of freedom is gapped with an optical gap between the upper and lower Hubbard bands $\Delta_{\rm opt} \simeq U - 4t + 8\ln(2)t^2/U$ \cite{ovchinnikov1992excitation}. Therefore, the optical conductivity is finite only above the optical gap at frequencies $\omega \simeq U \pm 4t$ at $T=0$ \cite{jeckelmann2000optical}. We do not expect this behaviour to change with SOC and magnetic field in the limit that $\lambda, b \ll t \ll U$.

If we ignore corrections of the order of $t/U$, there are no doubly occupied sites in the ground state. Using second order perturbation theory, an effective Hamiltonian can be written in terms of SU($2$) spin operators in the reduced Hilbert space generated by states without double occupancy, which reads
\begin{align}
\label{eq:spinham}
  \ham_{\text{spin}} =& \ham_{\text{ex}}  - \sum_l \vect{b} \cdot \vect{S}_l, \\
  \ham_{\text{ex}} =& \sum_l  J \left(	 S_l^x S_{l+1}^x + S_l^y S_{l+1}^y \right) + J^z S_l^z S_{l+1}^z \nonumber \\
  &+ \vect{D} \cdot \left( \vect{S}_l \times \vect{S}_{l+1} \right),
\end{align}
with 
\begin{equation}
  J = \frac{4(t^2 - \lambda^2)}{U}, \quad J^z = \frac{4(t^2 + \lambda^2)}{U}, \quad \vect{D} = \frac{8t\lambda}{U} \uvec{d}.
\end{equation}
The spin $z$-direction is defined by $S^z = \uvec{d} \cdot \vect{S}$.

At $T=0$, the current-current correlation function $\chi_{jj}(\omega)$, related to the optical conductivity by $\sigma'(\omega) =  {\rm Im}\{\chi_{jj}(\omega)\}/\omega$, is
\begin{align}
\label{eq:chiexact}
  \chi_{jj}(\omega) &= -\frac{1}{L} \mel{\psi_{0}}{j\frac{1}{E_0 - \ham + \omega + i\eta}j}{\psi_{0}} \nonumber \\ 
  					&= -\frac{1}{L} \sum_n \frac{\abs{\mel{\psi_{0}}{j}{\psi_{n}}}^2}{E_0 - E_n + \omega + i\eta},
\end{align}
where $\psi_{0}$ is the ground state of the original Hamiltonian (\ref{eq:tight}) with energy $E_0 = \mathcal{O}(t)$, and $\psi_{n}$ are the (optically allowed) excited state with energy $E_n = U + \mathcal{O}(t)$.

We are interested in the qualitative effect of SOC and magnetic field on the optical conductivity, and thus look for an effective expression of (\ref{eq:chiexact}) using the Hamiltonian (\ref{eq:spinham}). We also ignore the contribution of order $t/U$ in $E_0$ and $E_n$, so that $\sigma'(\omega)$ has only one contribution at $\omega = U$.
Let $\mathcal{P}_{\rm S}$ be the projection operator to the reduced Hilbert space. The ground state of (\ref{eq:spinham}) is then $\ket{\psi_{0}^{\rm s}} \equiv \mathcal{P}_{\text{S}} \ket{\psi_{0}}$.
The effective current-current correlation function in the $U \gg t$ limit is
\begin{equation}
\label{eq:correllargeU}
  \chi_{jj}(\omega) = -\frac{1}{L} \mel{\psi_{0}^{\rm s}}{j\frac{1}{\mathcal{O}(t) - [ U + \mathcal{O}(t)] + \omega + i\eta}j}{\psi_{0}^{\rm s}}.
\end{equation}
In the $U/t \rightarrow \infty$ limit, $\ket{\psi_{0}^{\rm s}} = \ket{\psi_0}$ and Eq.~(\ref{eq:correllargeU}) is exact. Neglecting the variation $\mathcal{O}(t)$ of the energy of the ground and excited states due to the dispersion of the bands, equation (\ref{eq:correllargeU}) factorizes and can be expressed in terms of spin operators using
\begin{equation}
   \frac{1}{U}\mathcal{P}_{\text{S}} j^2 \mathcal{P}_{\text{S}} = -\ham_{\text{ex}} + \text{const.}
\end{equation}
We naturally only find a resonance at the frequency $\omega = U$ and the optical conductivity reads
\begin{equation}
\label{eq:largUOC}
  \sigma'(\omega) = -\pi \left( \frac{1}{L} \langle \ham_{\text{ex}} \rangle_{\text{S}} - \frac{J^z}{4} \right) \delta(\omega - U),
\end{equation}
where $\langle \cdot \rangle_{\text{S}}$ refers to the statistical average using the full spin Hamiltonian (\ref{eq:spinham}) (the expression is also valid at finite temperature $T \ll \Delta_{\rm opt}$).

\begin{figure}[htb]
\centering
\includegraphics[width=0.46\textwidth]{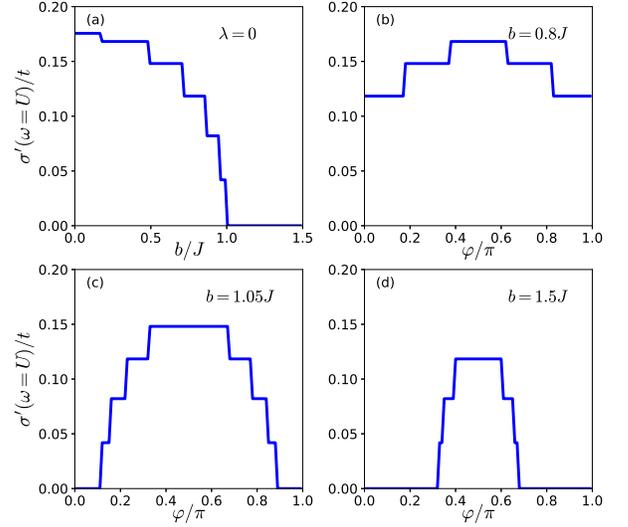}
\caption{Optical conductivity of the effective spin model (Eq.~\ref{eq:largUOC}) for $U/t = 50$ ($J/t = 0.08$) at $\omega=U$. The multiplicative factor in front of the Dirac delta function is plotted. (a): The amplitude is shown as a function of the magnetic field $b$ without SOC. (b), (c) and (d): The amplitude is shown as a function of $\varphi$, with SOC ($\lambda/t = 10^{-3}$), for $b = 0.8J$, $1.05J$ and $1.5J$, respectively. The calculations were made in a system of 12 spins at $T=0$.}
\label{fig:largeUresult}
\end{figure}

In Fig.~\ref{fig:largeUresult} we plot the optical conductivity calculated from Eq.~(\ref{eq:largUOC}) varring $\lambda$, $b$ and $\varphi$, in a system of $12$ sites at $T=0$. The step-like behaviour is due to this finite size effect. We see from Eq.~(\ref{eq:largUOC}) and Fig.~\ref{fig:largeUresult} (a) that the optical conductivity at $\omega = U$ vanishes for large $b$ in the absence of SOC. Indeed, for large values of $b$, the system becomes magnetically polarized and $\langle \ham_{\text{ex}} \rangle_{\text{S}}$ converges to $L J^z/4$. This is not surprising because the hopping is forbidden due to the Pauli principle when all the electrons have the same spin. However, SOC allows the electrons to rotate their spin while hopping. Therefore, the Pauli principle does not completely forbid the hopping and the optical conductivity recovers its finite value depending on $\varphi$, and is maximal when $\vect{b}$ and $\uvec{d}$ are perpendicular ($\varphi = \pi/2$), as seen in Fig.~\ref{fig:largeUresult}~(b), (c) and (d).\\

\section{Conclusion}
\label{sec:conclusion}

In this paper, we have investigated the synergetic effects of spin-orbit coupling and magnetic field in the 1D Hubbard model on the optical conductivity, in particular its dependence on the angle between the SOC vector and the magnetic field.

In the metallic phase (i.e. zero or exponentially small $\Delta_{\rm opt}$), due to SOC, the electric dipole spin resonance is possible and dominates over the purely magnetic resonance. We calculated the optical resonance for $U=0$ exactly. We described the resonance and observed characteristic dependences on the relative direction of the magnetic field and the SOC vector, and on the Fermi energy. Interestingly, a similar phenomenon has been observed experimentally in quasi-1D magnetic systems with Dzyaloshinskii-Moriya interaction, where the low-energy theory in terms of spinons is similar to the one for electrons in the metallic regime \cite{povarov2011modes,smirnov2015electron}. The measured magnetic dipole ESR signal similarly splits into distinct contributions from left and right movers, depending on the relative angle of the magnetic field. 

Then we described the evolution of the resonance for finite $U$ in small systems using exact diagonalization. We showed that at quarter-filling the Hubbard interaction does not destroy the original resonance but modifies its amplitude and frequency. This reflects the metallic behaviour of the system. In the half-filled case, the resonance observed at $U=0$ is enhanced for small $U$, where the optical gap is exponentially small, but then vanishes when $U$ is large enough, as the optical gap reaches $\Delta_{\rm opt} \gtrsim t$. The suppression of the EDSR can be understood as there are no gapless spinful particles which couple to the electric field. The magnetic susceptibility, however, is still finite and the magnetic dipole ESR signal does not vanish.

In the half-filling case, we also investigated the system from the strong coupling limit, $U/t \rightarrow \infty$. The charge degree of freedom is gapped and the spin resonance is not observed in the optical conductivity. In this case, the optical conductivity has finite contributions around $\omega \sim U \pm 4t$, which corresponds to the transitions between the upper and lower Hubbard bands, above the optical gap $\Delta_{\rm opt} \simeq U-4t$. We calculated the current-current response function neglecting all $\mathcal{O}(t/U)$ corrections, and showed its synergetic dependence on the external magnetic field and the spin-orbit coupling. 

\section*{Acknowledgements}

The authors thank Oleg Starykh for bringing our attention to Refs.~\cite{povarov2011modes,smirnov2015electron}.
A. B. acknowledges FMSP for the encouragement of the present study. 
H. K. was supported in part by JSPS KAKENHI 
Grant No. JP15K17719 and No. JP16H00985.
The present work was supported by the Elements Strategy Initiative Center for Magnetic Materials (ESICMM) under the outsourcing project of MEXT.
The numerical calculations were supported by the supercomputer center of ISSP of Tokyo University.

\bibliographystyle{apsrev4-1}

%

\end{document}